# Construction of Zero Cross Correlation Code using a type of Pascal's Triangle Matrix for Spectral Amplitude Coding Optical Code Division Multiple Access networks

K.S.Nisar

Department of Mathematics, Prince Sattam bin Abdulaziz University, Wadi Aldawaser, Riyadh Region, Kingdom of Saudi Arabia,

Email: ksnisar1@gmail.com, Fax.+9661588701, Phone:+966563456976

**Abstract:**

In this paper a new method to construct zero cross correlation code with the help of Pascal's triangle pattern called Pascal's Triangle Matrix Code (PTMC) for Spectral Amplitude Coding Optical Code Division Multiple Access (SAC-OCDMA) system is successfully developed. The advantages of this code are simplicity of code construction, flexibility of choosing code weight and number of users. The numerical comparison shows that the newly constructed code is better in code length than the existing codes such as Optical Orthogonal Code (OOC), Hadamard, Modified Frequency Hopping (MFH) and Modified Double Weight (MDW) codes and it supports more users than other Optical Spectrum Code Division Multiple Access (OSCDMA) codes.

## 1. Introduction:

The primary feature that differentiates Optical Code Division Multiple Access (OCDMA) from other multiple access techniques is that usage of orthogonal codes to permit multiple users to allow the same overlapping spectral range without interfering with each other. While the Multiple Access Interference (MAI) problems existing in optical code division multiple access (OCDMA) systems. In OCDMA system, due to the overlapping spectra of different users the phase induced intensity noise (PIIN) is sturdily related to multiple access interference (MAI) [1].The multiple access interference can be removed by balance detection scheme but the phase induced intensity noise from spontaneous emission of broad band source, is integrally remains. For this reason the researches carry attention to develop the code word in which the effect of MAI and PIIN of the total received power is reduced [2-6].

The basic multiple access techniques are Wave Division Multiple Access (WDMA), Time Division Multiple Access (TDMA) and Code Division Multiple Access (CDMA). The main limitations of TDMA and WDMA technologies are it allows multiple users to access a channel by allocating time slots and allocating wavelength or frequency to each user within each channel respectively. Also both technologies have a limited bandwidth for every user. The benefits of OCDMA systems are that the minimization of cross correlation to reduce a smaller value [7-10]. Although, the advantages of codes with zero cross correlation have less noise, which reduce the hardware complexity. Many of papers appeared in literature to study and analyze the performance of zero cross correlation code.

In [11] authors developed a new code structure for Spectral-Amplitude Coding Optical Code Division Multiple Access (OCDMA) system with zero cross-correlation. They showed that the zero cross correlation code that eliminate the PIIN and improve the system performance significantly.

The zero vectors combinatorial code family for spectral amplitude coding (SAC-OCDMA) based on combination of specific vectors with combinatorial theories is presented in [12].The proposed code have the advantages of flexibility of choosing the number of users and weights.

Garadi Ahmed and Ali Djebbari [13] proposed a new technique for constructing zero cross correlation codes (ZCC). To overcome the difficulties of code weight selection such as weight $W$ more than one, the weight is always fixed to the maximum number of users (i.e. the code size) authors constructed an efficient code word.

M.S.Anuar *et al*. [14] suggested a zero cross-correlation (ZCC) code to reduce the impact of system impairment and multiple access interference (MAI) in spectral amplitude coding optical code division multiple access (SAC–OCDMA) system. They showed that the established system not only preserves the capability of suppressing MAI, but also improves bit error rate performance as compared to conventional codes.

Recently K.S.Nisar [15] constructed the zero cross correlation code using a type of anti-diagonal identity column block matrices. The newly constructed zero cross correlation code showed that the code has better code length and it supports more users than other Optical Spectrum Code Division Multiple Access (OSCDMA) codes.

In this paper, we present a new zero cross correlation code with the help of Pascal's triangle called Pascal's Triangle Matrix Code (PTMC) to improve the performance of optical network. The advantages of the suggested new zero cross correlation codes are: i) any positive integer number of code weight ($W$ >1); ii) larger flexibility in choosing the number of users (free cardinality); iii) simple way of code construction using Pascal's type triangle pattern; and iv) cross correlation is equal to zero. The paper is structured as follows: The basic definitions of Pascal's triangle, Pascal's type triangle matrix, left Pascal's type and right Pascal's type matrix pattern, their properties are given in section 2. The code construction and code properties of the newly constructed codes are presented in Section 3. Comparison with reported codes and discussions are given in Section 4. Finally, conclusion is given in Section 5.

**2. Definitions and Properties.**

2.1 Pascal's Triangle

In mathematics, Pascal's triangle is a triangular array of the binomial coefficients. The rows of Pascal's triangle are conventionally enumerated starting with row $n = 0$ at the top. The entries in each row are numbered from the left beginning with $k = 0$ and are usually staggered relative to the numbers in the adjacent rows. The Pascal's triangle construction is associated to the binomial coefficients by Pascal's rule, which says that

$$\text{If} \quad (x+y)^n = \sum_{k=0}^{n} \binom{n}{k} x^{n-k} y^k \quad (1)$$

$$\text{then} \quad \binom{n}{k} = \binom{n-1}{k-1} + \binom{n-1}{k} \quad (2)$$

for any non-negative integer n and any integer $k$ between 0 and $n$.

The following pattern shows the first six rows of Pascal's triangle [16]

$$
\begin{array}{ccccccccc}
 & & & & 1 & & & & \\
 & & & 1 & & 1 & & & \\
 & & 1 & & 2 & & 1 & & \\
 & 1 & & 3 & & 3 & & 1 & \\
1 & & 4 & & 6 & & 4 & & 1 \\
\end{array}
$$
1   5   10   10   5   1

Now, we define the following

2.2 Pascal's type triangle Matrix (PTM)

By considering the binary numbers, the Pascal's type triangle Matrix starts from second row of Pascal's triangle constructed as follows; *i)* replace all the numbers other than 1 by 0, *ii)* Fill the outer branches of the triangle by 0. A $4\times 8$ Pascal's type matrix starting from second row of Pascal's triangle as follows

$$
(PTM)_2 = \begin{bmatrix} 0 & 0 & 0 & 1 & 1 & 0 & 0 & 0 \\ 0 & 0 & 1 & 0 & 0 & 1 & 0 & 0 \\ 0 & 1 & 0 & 0 & 0 & 0 & 1 & 0 \\ 1 & 0 & 0 & 0 & 0 & 0 & 0 & 1 \end{bmatrix}
$$

A $4\times 9$ Pascal's type matrix starting from third row of Pascal's triangle as follows

$$
(PTM)_3 = \begin{bmatrix} 0 & 0 & 0 & 1 & 1 & 1 & 0 & 0 & 0 \\ 0 & 0 & 1 & 0 & 0 & 0 & 1 & 0 & 0 \\ 0 & 1 & 0 & 0 & 0 & 0 & 0 & 1 & 0 \\ 1 & 0 & 0 & 0 & 0 & 0 & 0 & 0 & 1 \end{bmatrix}
$$

Similarly a general Pascal's type triangle matrix is defined as

$$
(PTM)_n = \begin{bmatrix} 0 & 0 & \cdots & 0 & \overbrace{1 & \cdots & \cdots & 1}^{n\text{-times}} & 0 & \cdots & 0 & 0 \\ \vdots & \vdots & 0 & 1 & 0 & \cdots & \cdots & 0 & 1 & 0 & \vdots & \vdots \\ \vdots & 0 & 1 & 0 & \cdots & \cdots & \cdots & \cdots & 0 & 1 & 0 & \vdots \\ 0 & 1 & 0 & \cdots & \cdots & \cdots & \cdots & \cdots & \cdots & 0 & 1 & 0 \\ 1 & 0 & \cdots & \cdots & \cdots & \cdots & \cdots & \cdots & \cdots & \cdots & 0 & 1 \end{bmatrix}
$$

2.3 Left Pascal's type triangle Matrix (LPTM)

The LPTM defined by considering the left half of Pascal's triangle matrix. A 4x3 LPTM defined as follows

$$\begin{bmatrix} 0 & 0 & 0 \\ 0 & 0 & 1 \\ 0 & 1 & 0 \\ 1 & 0 & 0 \end{bmatrix}$$

2.4 Right Pascal's type triangle Matrix (RPTM)

The RPTM defined by considering the right half of Pascal's triangle matrix. A 4x3 RPTM defined as follows

$$\begin{bmatrix} 0 & 0 & 0 \\ 1 & 0 & 0 \\ 0 & 1 & 0 \\ 0 & 0 & 1 \end{bmatrix}$$

**3. Code constructions and Examples.**

**Steps of constructions:**

Step 1: Fill $W$ times '1s' in the $W^{th}$ row of Pascal's triangle matrix PTM and add '1s' diagonally

Step 2: Fill the remaining places with '0s' in the PTM matrix.

Step 3: If the code weight $W = 2n+1, n = 1,2,3\ldots$ that is $W$ is odd, then add $W-(n+1)$ LTM and $W-(n+2)$ RTM with Pascal's triangle matrix formed using step 1 and step 2.

Step 4: If the code weight $W > 2$ and $W = 2n, n = 2,3\ldots$ that is $W$ is even, then add $W-(n+1)$ LTM and $W-(n+1)$ RTM with Pascal's Triangle matrix obtained by step 1 and step 2

Step 5: To increase the number of users, increase the size of LTM and RTM and repeat step 1-step4

**Case 1. $W=2$ and $N=4$ ($W$ is even).**

The case of $W=2$ the code pattern directly generated from the Pascal's type triangle matrix shown in Table 1.

**Case 2. W=4, N=4 (W is even and W >2 )**

Step 1: Fill $W = 4$ times '1s' in the 4$^{th}$ row of Pascal's triangle matrix (PTM) and add '1s' diagonally

Step 2: Fill the remaining places with '0s' in the Pascal's type triangle matrix.

Step 3: Here the code weight $W>2$ and $w=2n=2\times 2=4$ is even, then we have to add $W-(n+1)=4-(2+1)=1$ LTM and $W-(n+1)=4-(2+1)=1$ RTM with Pascal's Triangle matrix obtained by step 1 and step 2 (see Table 2).

**Case 3 . W=3, N=4 (W is odd)**

Step 1: Fill $W = 3$ times '1s' in the 3$^{rd}$ row of Pascal's triangle matrix (PTM) and add '1s' diagonally

Step 2: Fill the remaining places with '0s' in the Pascal's type triangle matrix.

Step 3: Here the code weight $w=2n+1=2\times 1+1=3$ is odd, then we have to add $W-(n+1)=3-(1+1)=1$ LTM and $W-(n+2)=3-(1+2)=0$ RTM with Pascal's Triangle matrix obtained by step 1 and step 2 (see Table 3).

To increase the code weight, we consider the following examples

For **W=6, N=4**

Here the code weight $W>2$ and $w=2n=2\times 3=6$ is even, then we have to add $W-(n+1)=6-(3+1)=2$ LTM and $W-(n+1)=6-(3+1)=2$ RTM with Pascal's triangle matrix obtained by step 1 and step 2 (see Table 4).

For **W =8, N=4**

Here the code weight $W>2$ and $w=2n=2\times 4=8$ is even, then we have to add $W-(n+1)=8-(4+1)=3$ LTM and $W-(n+1)=8-(4+1)=3$ RTM with Pascal's triangle matrix obtained by step 1 and step 2 (see Table 5).

For *W=5, N=4*

Here the code weight $w = 2n+1 = 2\times 2+1 = 5$ is odd, then we have to add $W-(n+1) = 5-(2+1) = 2$ LTM and $W-(n+2) = 5-(2+2) = 1$ RTM with Pascal's Triangle matrix obtained by step 1 and step 2 (see Table 6).

*For W=7, N=4*

Here the code weight $w = 2n+1 = 2\times 3+1 = 7$ is odd, then we have to add $W-(n+1) = 7-(3+1) = 3$ LTM and $W-(n+2) = 7-(3+2) = 1$ RTM with Pascal's Triangle matrix obtained by step 1 and step 2 (see Table 7).

To increase the number of users we have to add the number of rows of the PT matrix as follows

For *W=2, N=5*, the number of rows can increase symmetrically without any mathematical difficulties (see Table 8)

For *W=4, N=6*

In this case the code weight $W > 2$ and $w = 2n = 2\times 2 = 4$ is even, then we have to add $W-(n+1) = 4-(2+1) = 1$ LTM and $W-(n+1) = 4-(2+1) = 1$ RTM with Pascal's triangle matrix obtained by step 1 and step 2 (see Table 9).

For *W= 3, N=5*

Here the code weight $w = 2n+1 = 2\times 1+1 = 3$ is odd, then we have to add $W-(n+1) = 3-(1+1) = 1$ LTM and $W-(n+2) = 3-(1+2) = 0$ RTM with Pascal's Triangle matrix obtained by step 1 and step 2 (see Table 10).

For *W=3,N=6*

In this case the code weight $w = 2n+1 = 2\times1+1 = 3$ is odd, then we have to add $W-(n+1) = 3-(1+1) = 1$ LTM and $W-(n+2) = 3-(1+2) = 0$ RTM with Pascal's Triangle matrix obtained by step 1 and step 2 (see Table 11).

**4. Code evaluation and Comparison**

Many codes have been proposed for OSCDMA such as Optical Orthogonal Code (OOC) [17], Modified Frequency Hoping (MFH) code and Modified double weight (MDW) code [1]. However there are many limitations in these codes such as complication on code construction (example MFH and OOC codes) ,the cross correlation not ideal or length of the code is too long (example. OOC). The code length of newly developed code PTMC as same as ZCC [11], but the complication of solving equation (7) of [11] is relaxed by PTMC code.

In Table 12, PTMC shows flexibility in terms of choosing the code weight and the number of users due to the use of a novel technique in code construction, which generates many possibilities from a given number of users and weights.

**5 Conclusions:**

In this work, a new zero cross correlation code for SAC-OCDMA system called Pascal's triangle matrix code (PTMC) is defined and developed successfully. By comparing to other existing codes in the SAC-OCDMA system ,Pascal's triangle code have the advantages that is the easiest way of code construction with the help of newly defined Pascal's type triangle (including Pascal's left triangle and Pascal's right triangle) pattern. The strongest contribution of PTMC is the elimination of PIIN as it has zero cross correlation between the codes. PTMC has numerous advantages such as flexibility of choosing number of users and the code weight. The Pascal's triangle code better in length and the code can construct easily without any mathematical complication.

Table 1. PTM code patterns for $W=2$ and $N=4$

| 0 | 0 | 0 | 0 | 1 | 1 | 0 | 0 | 0 | 0 |
|---|---|---|---|---|---|---|---|---|---|
| 0 | 0 | 0 | 1 | 0 | 0 | 1 | 0 | 0 | 0 |
| 0 | 0 | 1 | 0 | 0 | 0 | 0 | 1 | 0 | 0 |
| 0 | 1 | 0 | 0 | 0 | 0 | 0 | 0 | 1 | 0 |

Table 2. PTM code patterns for $W=4$ and $N=4$

| 0 | 0 | 0 | 0 | 0 | 0 | 1 | 1 | 1 | 1 | 0 | 0 | 0 | 0 | 0 | 0 |
|---|---|---|---|---|---|---|---|---|---|---|---|---|---|---|---|
| 0 | 0 | 1 | 0 | 0 | 1 | 0 | 0 | 0 | 0 | 1 | 0 | 0 | 1 | 0 | 0 |
| 0 | 1 | 0 | 0 | 1 | 0 | 0 | 0 | 0 | 0 | 0 | 1 | 0 | 0 | 1 | 0 |
| 1 | 0 | 0 | 1 | 0 | 0 | 0 | 0 | 0 | 0 | 0 | 0 | 1 | 0 | 0 | 1 |

Table 3. PTM code patterns for $W=3$ and $N=4$

| 0 | 0 | 0 | 0 | 0 | 0 | 1 | 1 | 1 | 0 | 0 | 0 |
|---|---|---|---|---|---|---|---|---|---|---|---|
| 0 | 0 | 1 | 0 | 0 | 1 | 0 | 0 | 0 | 1 | 0 | 0 |
| 0 | 1 | 0 | 0 | 1 | 0 | 0 | 0 | 0 | 0 | 1 | 0 |
| 1 | 0 | 0 | 1 | 0 | 0 | 0 | 0 | 0 | 0 | 0 | 1 |

Table 4. PTM code patterns for $W=6$ and $N=4$

| 0 | 0 | 0 | 0 | 0 | 0 | 0 | 0 | 0 | 1 | 1 | 1 | 1 | 1 | 1 | 0 | 0 | 0 | 0 | 0 | 0 | 0 | 0 |
|---|---|---|---|---|---|---|---|---|---|---|---|---|---|---|---|---|---|---|---|---|---|---|
| 0 | 0 | 1 | 0 | 0 | 1 | 0 | 0 | 1 | 0 | 0 | 0 | 0 | 0 | 0 | 1 | 0 | 0 | 1 | 0 | 0 | 1 | 0 | 0 |
| 0 | 1 | 0 | 0 | 1 | 0 | 0 | 1 | 0 | 0 | 0 | 0 | 0 | 0 | 0 | 1 | 0 | 0 | 1 | 0 | 0 | 1 | 0 |
| 1 | 0 | 0 | 1 | 0 | 0 | 1 | 0 | 0 | 0 | 0 | 0 | 0 | 0 | 0 | 0 | 1 | 0 | 0 | 1 | 0 | 0 | 1 |

Table 5. PTM code patterns for $W=8$ and $N=4$

| 0 | 0 | 0 | 0 | 0 | 0 | 0 | 0 | 0 | 0 | 0 | 0 | 1 | 1 | 1 | 1 | 1 | 1 | 1 | 1 | 0 | 0 | 0 | 0 | 0 | 0 | 0 | 0 | 0 | 0 |
|---|---|---|---|---|---|---|---|---|---|---|---|---|---|---|---|---|---|---|---|---|---|---|---|---|---|---|---|---|---|
| 0 | 0 | 1 | 0 | 0 | 1 | 0 | 0 | 1 | 0 | 0 | 1 | 0 | 0 | 0 | 0 | 0 | 0 | 0 | 0 | 1 | 0 | 0 | 1 | 0 | 0 | 1 | 0 | 0 | 1 | 0 | 0 |
| 0 | 1 | 0 | 0 | 1 | 0 | 0 | 1 | 0 | 0 | 1 | 0 | 0 | 0 | 0 | 0 | 0 | 0 | 0 | 0 | 0 | 1 | 0 | 0 | 1 | 0 | 0 | 1 | 0 | 0 | 1 | 0 |
| 1 | 0 | 0 | 1 | 0 | 0 | 1 | 0 | 0 | 1 | 0 | 0 | 0 | 0 | 0 | 0 | 0 | 0 | 0 | 0 | 1 | 1 | 0 | 0 | 0 | 0 | 1 | 0 | 0 | 1 |

Table 6. PTM code patterns for *W*=5 and *N*=4

| 0 | 0 | 0 | 0 | 0 | 0 | 0 | 0 | 0 | 1 | 1 | 1 | 1 | 1 | 0 | 0 | 0 | 0 | 0 | 0 |
| 0 | 0 | 1 | 0 | 0 | 1 | 0 | 0 | 1 | 0 | 0 | 0 | 0 | 0 | 1 | 0 | 0 | 1 | 0 | 0 |
| 0 | 1 | 0 | 0 | 1 | 0 | 0 | 1 | 0 | 0 | 0 | 0 | 0 | 0 | 0 | 1 | 0 | 0 | 1 | 0 |
| 1 | 0 | 0 | 1 | 0 | 0 | 1 | 0 | 0 | 0 | 0 | 0 | 0 | 0 | 0 | 0 | 1 | 0 | 0 | 1 |

Table 7. PTM code patterns for *W*=7 and *N*=4

| 0 | 0 | 0 | 0 | 0 | 0 | 0 | 0 | 0 | 0 | 0 | 0 | 1 | 1 | 1 | 1 | 1 | 1 | 1 | 0 | 0 | 0 | 0 | 0 | 0 | 0 | 0 |
| 0 | 0 | 1 | 0 | 0 | 1 | 0 | 0 | 1 | 0 | 0 | 1 | 0 | 0 | 0 | 0 | 0 | 0 | 0 | 1 | 0 | 0 | 1 | 0 | 0 | 1 | 0 | 0 |
| 0 | 1 | 0 | 0 | 1 | 0 | 0 | 1 | 0 | 0 | 1 | 0 | 0 | 0 | 0 | 0 | 0 | 0 | 0 | 0 | 1 | 0 | 0 | 1 | 0 | 0 | 1 | 0 |
| 1 | 0 | 0 | 1 | 0 | 0 | 1 | 0 | 0 | 1 | 0 | 0 | 0 | 0 | 0 | 0 | 0 | 0 | 0 | 0 | 0 | 1 | 0 | 0 | 1 | 0 | 0 | 1 |

Table 8. PTM code patterns for *W*=2 and *N*=5

| 0 | 0 | 0 | 0 | 1 | 1 | 0 | 0 | 0 | 0 |
| 0 | 0 | 0 | 1 | 0 | 0 | 1 | 0 | 0 | 0 |
| 0 | 0 | 1 | 0 | 0 | 0 | 0 | 1 | 0 | 0 |
| 0 | 1 | 0 | 0 | 0 | 0 | 0 | 0 | 1 | 0 |
| 1 | 0 | 0 | 0 | 0 | 0 | 0 | 0 | 0 | 1 |

Table 9. PTM code patterns for *W*=4 and *N*=6

| 0 | 0 | 0 | 0 | 0 | 0 | 0 | 0 | 0 | 0 | 1 | 1 | 1 | 1 | 0 | 0 | 0 | 0 | 0 | 0 | 0 | 0 | 0 |
| 0 | 0 | 0 | 0 | 1 | 0 | 0 | 0 | 0 | 1 | 0 | 0 | 0 | 0 | 1 | 0 | 0 | 0 | 0 | 1 | 0 | 0 | 0 | 0 |
| 0 | 0 | 0 | 1 | 0 | 0 | 0 | 0 | 1 | 0 | 0 | 0 | 0 | 0 | 0 | 1 | 0 | 0 | 0 | 0 | 1 | 0 | 0 | 0 |
| 0 | 0 | 1 | 0 | 0 | 0 | 0 | 1 | 0 | 0 | 0 | 0 | 0 | 0 | 0 | 0 | 1 | 0 | 0 | 0 | 0 | 1 | 0 | 0 |
| 0 | 1 | 0 | 0 | 0 | 0 | 1 | 0 | 0 | 0 | 0 | 0 | 0 | 0 | 0 | 0 | 0 | 1 | 0 | 0 | 0 | 0 | 1 | 0 |
| 1 | 0 | 0 | 0 | 0 | 1 | 0 | 0 | 0 | 0 | 0 | 0 | 0 | 0 | 0 | 0 | 0 | 0 | 1 | 0 | 0 | 0 | 0 | 1 |

Table 10. PTM code patterns for *W*=3 and *N*=5

| 0 | 0 | 0 | 0 | 0 | 0 | 0 | 1 | 1 | 1 | 0 | 0 | 0 | 0 |
| 0 | 0 | 0 | 0 | 0 | 0 | 1 | 0 | 0 | 0 | 1 | 0 | 0 | 0 |
| 0 | 0 | 1 | 0 | 0 | 1 | 0 | 0 | 0 | 0 | 0 | 1 | 0 | 0 |
| 0 | 1 | 0 | 0 | 1 | 0 | 0 | 0 | 0 | 0 | 0 | 0 | 1 | 0 |
| 1 | 0 | 0 | 1 | 0 | 0 | 0 | 0 | 0 | 0 | 0 | 0 | 0 | 1 |

Table 11. P TM code patterns for *W*=3 and *N*=6

| 0 | 0 | 0 | 0 | 0 | 0 | 0 | 0 | 0 | 0 | 1 | 1 | 1 | 0 | 0 | 0 | 0 | 0 |
| 0 | 0 | 0 | 0 | 1 | 0 | 0 | 0 | 0 | 1 | 0 | 0 | 0 | 1 | 0 | 0 | 0 | 0 |
| 0 | 0 | 0 | 1 | 0 | 0 | 0 | 0 | 1 | 0 | 0 | 0 | 0 | 0 | 1 | 0 | 0 | 0 |
| 0 | 0 | 1 | 0 | 0 | 0 | 0 | 1 | 0 | 0 | 0 | 0 | 0 | 0 | 0 | 1 | 0 | 0 |
| 0 | 1 | 0 | 0 | 0 | 0 | 1 | 0 | 0 | 0 | 0 | 0 | 0 | 0 | 0 | 0 | 1 | 0 |
| 1 | 0 | 0 | 0 | 0 | 1 | 0 | 0 | 0 | 0 | 0 | 0 | 0 | 0 | 0 | 0 | 0 | 1 |

Table 12. SAC-OCDMA code comparison

| Code | Existence | Code weight | Cross Correlation | Code length |
|---|---|---|---|---|
| MFH | $K = Q^2$ | $Q+1$ | $\lambda=1$ | $N=Q^2+Q$ |
| MDW | $K = n$ | Even | $\lambda=1$ | $N = 3n + \frac{8}{3}\left[\sin\left(\frac{k\pi}{3}\right)\right]^2$ |
| ZCC [11] | $K=2^m$ | $2^{m-1}$ | $\lambda=0$ | $C=2^m$ |
| Hadamard[18] $M \geq 2$ | $K= 2^{M-1}$ | $2^{M-1}$ | $\lambda=2^{M-2}$ | $N=2^M$ |
| PTMC | Any number | For natural number $\geq 2$ | $\lambda=0$ | $L = N \times W$ |